\documentclass{PoS}

\usepackage{graphicx}
\usepackage{epsfig}         
\usepackage[tight,hang,raggedright]{subfigure} 
\usepackage{picins}                     
\usepackage{latexsym}

\newcommand{\mpp}{m_{\pi\pi}}

\newcommand{\Li}{\mathrm{Li}}

\title{Electroproduction of pion pairs}

\ShortTitle{Di-pion electroproduction}

\author{\speaker{D.Yu. Ivanov}\thanks{D.I. thanks the organizers of the workshop for their warm
hospitality. The work of D.I. was partially supported by grants
RFBR-06-02-16064-a, NSh-5362.2006.2. This work is
supported by the Helmholtz Association, contract number
VH-NG-004.}
         \\
        Sobolev Institute of Mathematics, 630090 Novosibirsk, Russia\\
        E-mail: \email{d-ivanov@math.nsc.ru}}

\author{M. Diehl\\
        Theory Group, Deutsches Elektronen Synchrotron DESY, 22603 Hamburg, Germany \\
        E-mail: \email{mdiehl@mail.desy.de}}

\author{A. Sch\"afer\\
        Institut f\"ur Theoretische Physik, Universitaet Regensburg,
93040 Regensburg, Germany\\
        E-mail: \email{andreas.schaefer@physik.uni-regensburg.de}}

\author{N. Warkentin\\
        Institut f\"ur Theoretische Physik, Universitaet Regensburg,
93040 Regensburg, Germany\\
        E-mail: \email{nikolaus.warkentin@physik.uni-regensburg.de}}

\abstract{We study hard exclusive electroproduction of two pions in
the QCD factorization approach at next-to-leading order (NLO) in
the strong coupling. The pion
pair can be produced both in an isovector and in  an isoscalar
state. The angular distribution of the produced pion pairs allows
one to
project out a component that depends on the interference of the
isovector and isoscalar channels. Using specific models for the involved
 generalized parton distributions and
two-pion distribution amplitudes we investigate the
angular distributions of the pion pair in NLO and compare them with
HERMES data. The differences between the LO and NLO results are
moderate and the agreement with data is satisfactory, though not
perfect. Expecting new results from COMPASS collaboration we perform
the calculation also for the COMPASS kinematics.}

\FullConference{DIFFRACTION 2006 -
International Workshop on Diffraction in High-Energy Physics \\
         September 5-10 2006\\
         Adamantas, Milos island, Greece}

\begin{document}

\section{Introduction}

In the last years it has been realized that a
large class of hard exclusive reactions
can be treated in QCD factorization framework, absorbing all
non-perturbative soft physics in suitable
generalized parton distributions (GPDs) and distribution
      amplitudes (DAs).
GPDs encode valuable information about hadron structure,
including some aspects which cannot be deduced directly from
experiment, like the transverse spatial distribution of partons and
the total angular momentum. For more details we refer to the reviews
\cite{Goeke:2001tz,Diehl:2003ny,Belitsky:2005qn}.

The large amount of information contained in GPDs implies
that much and diverse data is needed to
determine their functional form in the three variables on
     which they depend. One of the channels
in which data is already
available is the exclusive electroproduction of pion pairs \cite{Airapetian:2004sy}.
This was already studied by some of us
\cite{Lehmann-Dronke:1999aq,Lehmann-Dronke:2000xq} in the past in LO
and is now analyzed in NLO.

\section{The amplitude of the process}

The analysis of the di-pion electroproduction is reduced essentially to
that one for the subprocess
 \begin{equation}
    \gamma_L^*(q) + N(p)\rightarrow \pi^a(k^a) +\pi^b(k^b)+N(p^\prime),
    \label{subprc}
  \end{equation}
where a longitudinally polarized virtual photon $\gamma_L^*$ with
momentum $q$ hits a nucleon $N$ with momentum $p$ and  produces a
final state nucleon  $N$
with momentum $p^\prime$ and two pions with momenta
$k^a$ and $k^b$.
 We use the conventional variables
\begin{equation}
q^2=-Q^2\, , \quad \Delta=p^\prime - p\, , \quad \quad \Delta^2=t \, ,
\quad x_{\rm{Bj}}=\frac{Q^2}{2p\cdot q}\, ,
\end{equation}
and denote by $m_\pi, m_N$ and $m_{\pi\pi}$ the pion, the nucleon and the
di-pion mass, $m^2_{\pi\pi}=(k^a+k^b)^2$. We consider the limit
that virtuality of the photon is large,
\begin{equation}
 Q^2\gg |t| , m_N^2  , m^2_\pi , m^2_{\pi\pi} \,\,  ,
\end{equation}
 in fact, $Q^2$ is not larger than the energy scale $W^2$.
In this case the amplitude, according to the factorization theorem
in
\cite{Collins:1996fb}, may be written as a convolution of
hard coefficient function and soft parts, parameterized by GPDs,
and 2$\pi$DAs.
Note that the subprocess with longitudinally
polarized photon gives the leading contribution to the
reaction, the contribution of transverse
polarization is suppressed by a power $1/Q$.

At leading twist the pion pair can be produced both in
isoscalar and isovector states
\begin{equation}
    T^{\pi^+\pi^-}=T^{I=0}+T^{I=1} \, ,
\end{equation}
where
{
 \begin{eqnarray}
  \label{I1}
  T^{I=1}&= & \frac{2\pi\sqrt{4\pi \alpha}}{N_c Q\xi}
  \int\limits^1_{-1}d\tau
  \int\limits^1_{0}dz \, \Phi^{I=1}(z,\zeta,m_{\pi\pi})
  \sum_{f=u,d}e_f\tau_3^{f}\left[
  Q^{(+)}(z,\tau)F^{f(+)} (\tau,\xi,t)\phantom{\sum_f}\right.
   \\
  &&
  \left.
  + G^{(+)}(z,\tau)F^{g} (\tau,\xi,t)
  +R^{(+)}(z,\tau)
  \sum_{f^\prime=u,d,s} F^{f^\prime (+)} (\tau,\xi,t)
  \right] \, , \nonumber \\
  T^{I=0}&= &\frac{2\pi\sqrt{4\pi \alpha}}{N_c Q\xi}
  \int\limits^1_{-1}d\tau
  \int\limits^1_{0}dz
  \sum_{f=u,d}e_f F^{f(-)}(\tau,\xi,t)
  \left[ Q^{(-)}(z,\tau)\Phi^{I=0}_f(z,\zeta,m_{\pi\pi})\right.
  \nonumber \\
  &&
  \left.+
  2\xi G^{(-)}(z,\tau)\Phi^G(z,\zeta,m_{\pi\pi})
  +\sum_{f^\prime=u,d,s}R^{(-)}(z,\tau)\Phi^{I=0}_{f^\prime}(z,\zeta,m_{\pi\pi})
  \right]\, .\nonumber
  \end{eqnarray}
Here
$\tau_3^u=1$ and $\tau_3^d=-1$ for the up- and down-quark,
$\xi=x_{\rm{Bj}}/(2-x_{\rm{Bj}})$
is the skewness parameter, $N_c$ the number of colors
      and $\alpha$ the fine structure constant.  $Q^{(\pm)}$, $G^{(\pm)}$
      and $R^{(\pm)}$ denote hard coefficient functions, whereas $F$ and
      $\Phi$ stand for
 GPDs and 2$\pi$DAs.
The
integration in (\ref{I1}) runs over parton momentum fractions $\tau$ and $z$.
\begin{figure}
\begin{center}
\includegraphics[scale=0.22]{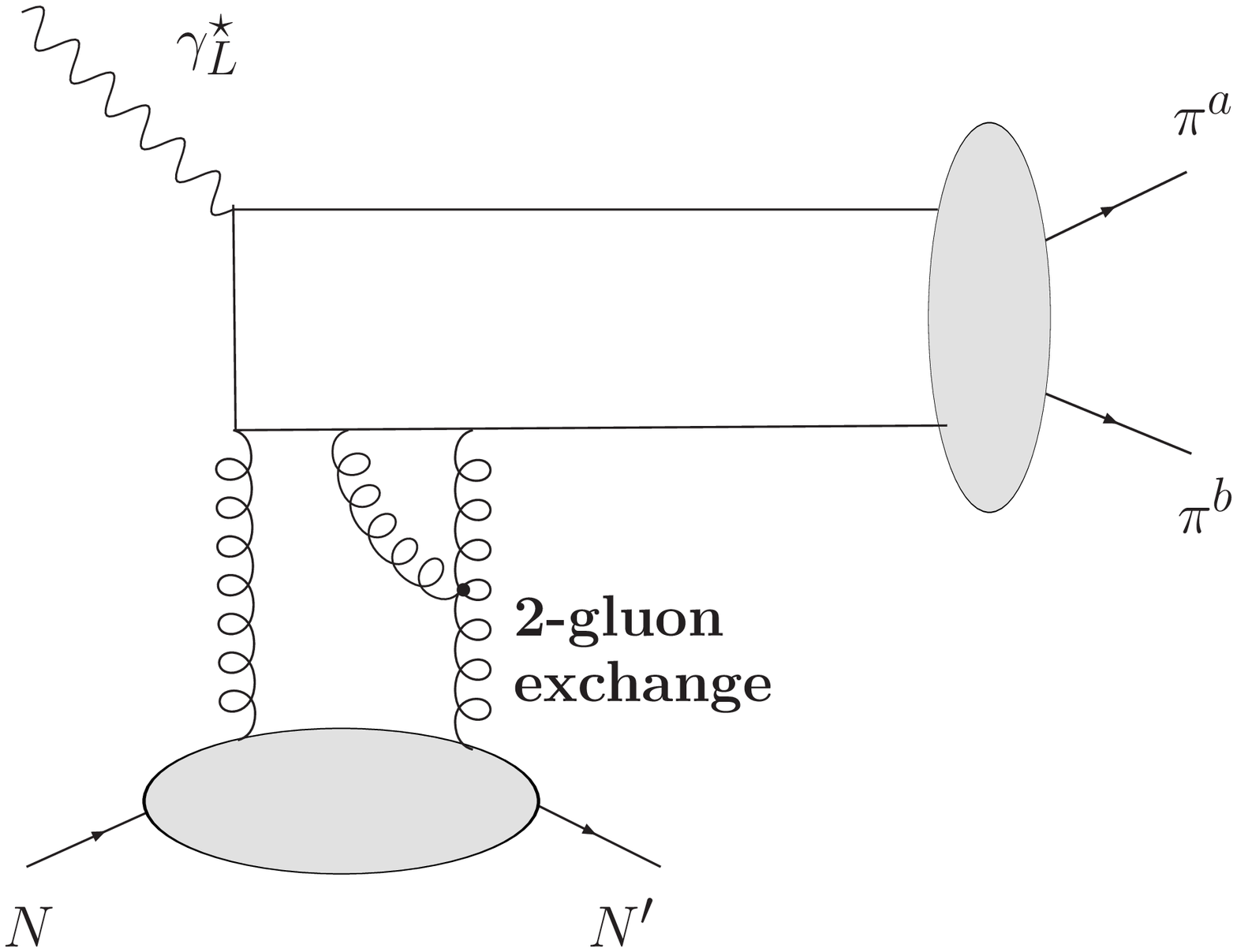}
\includegraphics[scale=0.22]{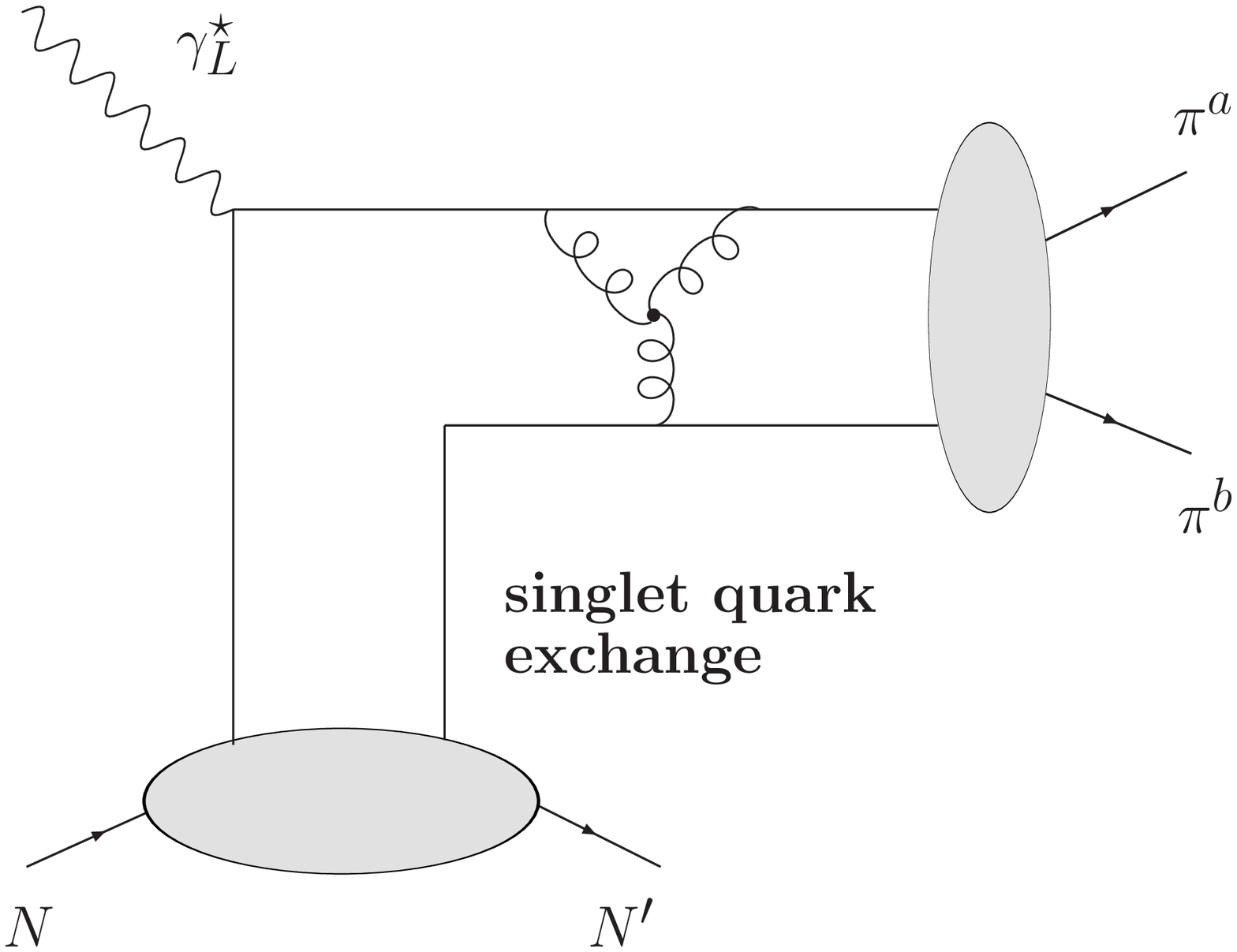}
\includegraphics[scale=0.22]{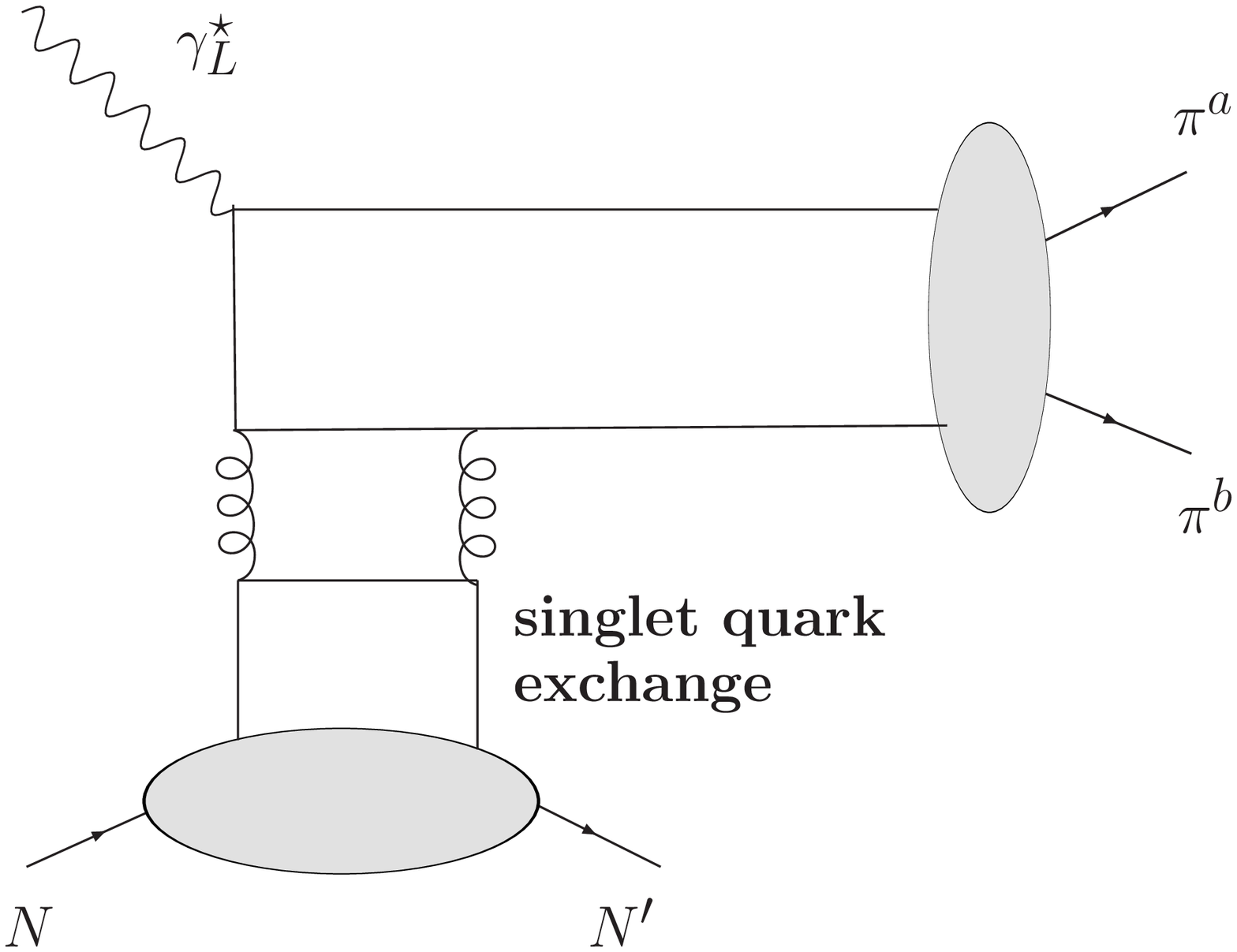}
\end{center}
\begin{center}
\includegraphics[scale=0.22]{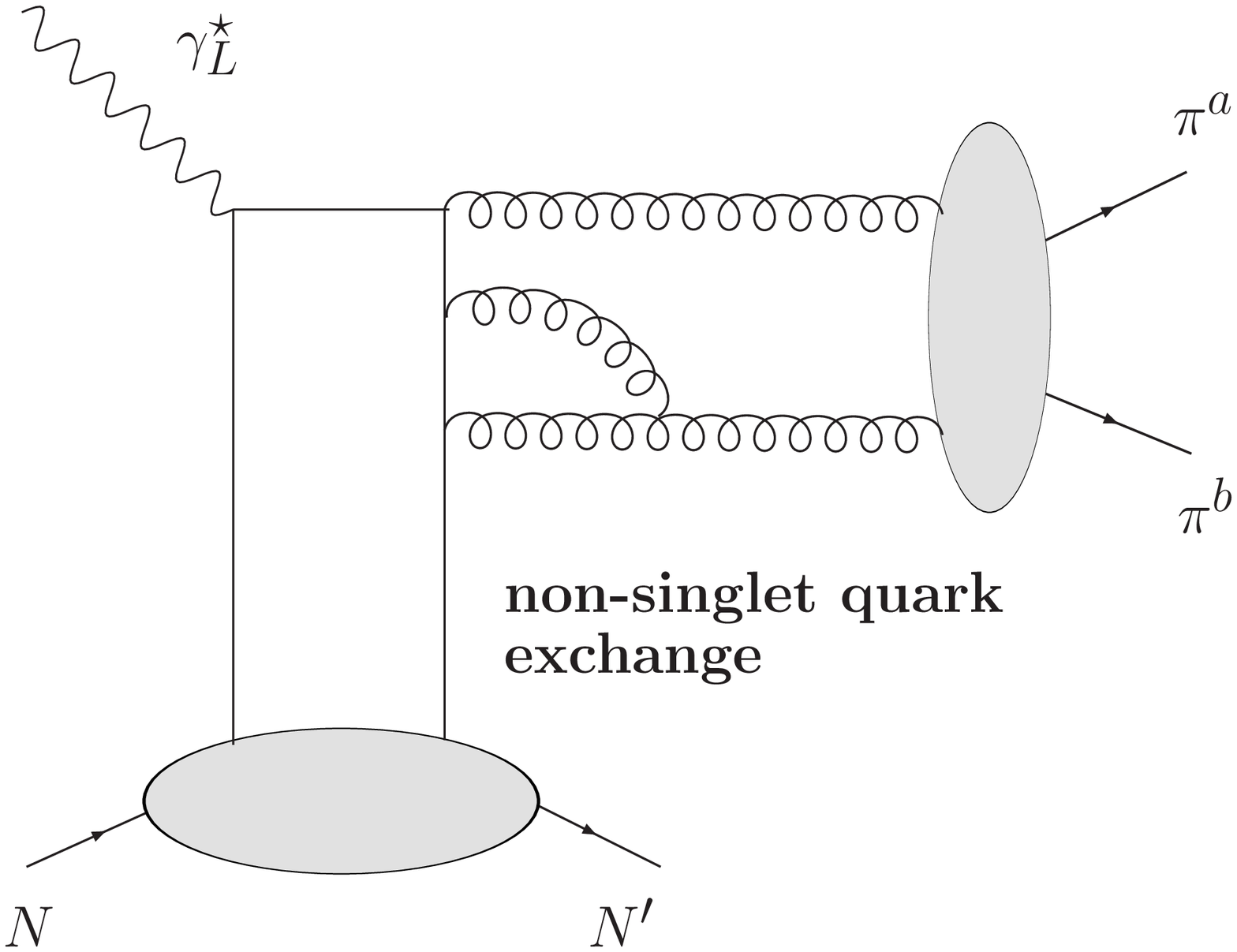}
\includegraphics[scale=0.22]{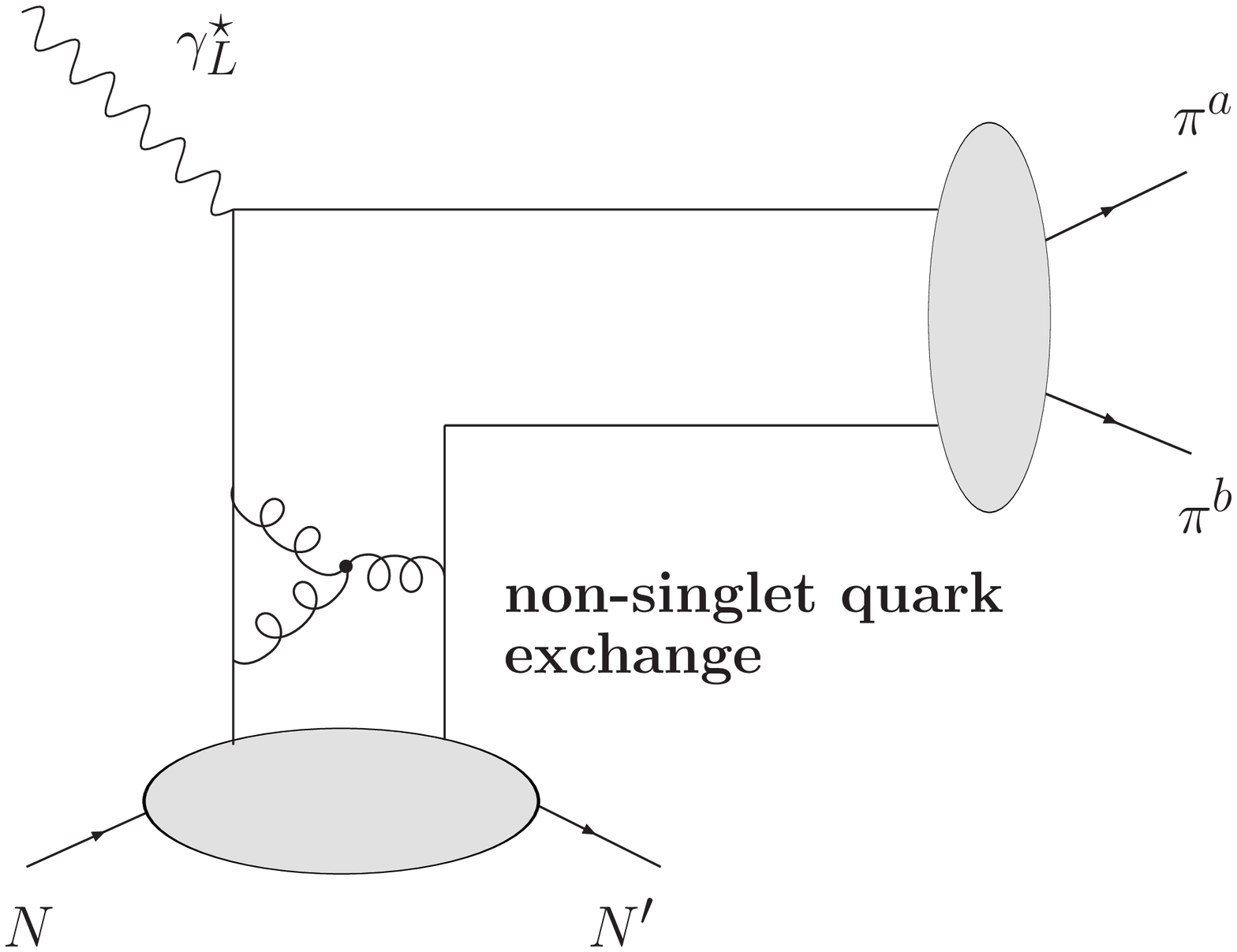}
\includegraphics[scale=0.22]{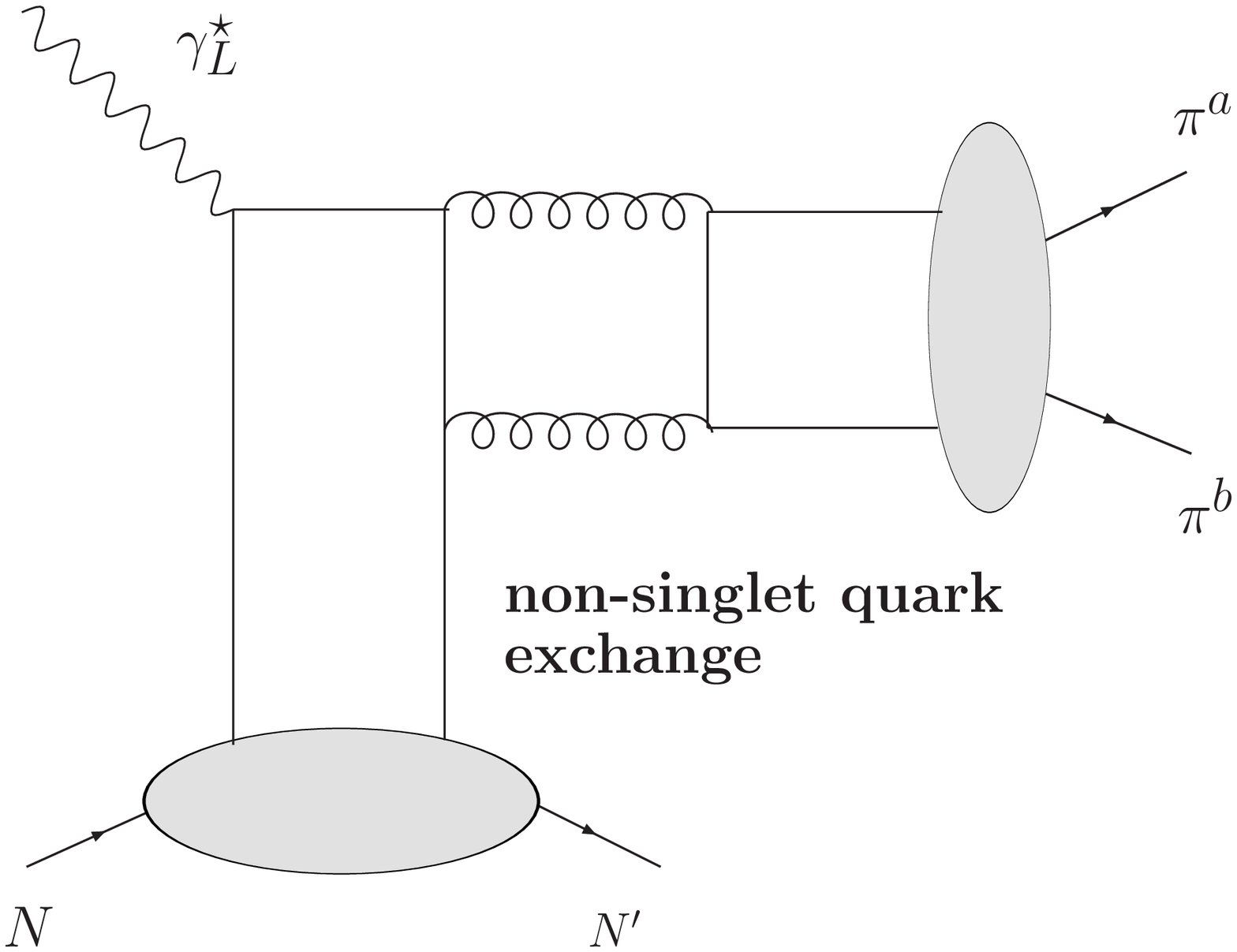}
\end{center}
\caption{Typical NLO diagrams for di-pion production in the
isovector (upper line) and isoscalar (lower line) state. The diagrams
related to each other by crossing are displayed one upon another.}
\label{f1}
\end{figure}
Due to charge conjugation invariance the production of an
      isovector pion pair, described by a $C$ odd quark DA, is
      mediated by gluon and $C$ even quark GPDs, $F^g$ and $F^{(+)}$.
      In contrast, the $C$ odd quark GPD $F^{(-)}$ contributes for the
      production of an isoscalar pion pair, which is produced by
      hadronization of a a gluon pair or a $C$ even quark pair.

The coefficient functions
represent the amplitudes for the scattering of collinear
   partons.
Typical NLO diagrams
are shown in Fig.~\ref{f1}, where the graphs in the first, the second and the third
columns contribute to  $R^{(\pm)}, G^{(\pm)}$ and $Q^{(\pm)}$ respectively.
$R^{(+)}, G^{(+)},Q^{(+)}$ were calculated in
the $\overline{\textrm{MS}}$ scheme
in \cite{Ivanov:2004zv}, where electroproduction of light vector
mesons was studied at NLO. For example
\begin{eqnarray}
&
R^{(+)}(z,\tau)=
 \alpha_s^2(\mu_R^2) C_F \frac{\displaystyle 1}{\displaystyle 8\pi z \bar{z}}
{\cal
R}\left(z,\frac{\tau-\xi}{2\xi}\right) \, ,
& \\
& {\cal R}(z,y)=\Biggl\{
\frac{2y+1}{y(y+1)}\left[\frac{y}{2}\ln^2(-y)-
\frac{y+1}{2}\ln^2(y+1)
 +
\left(\ln\left(\frac{Q^2z}{\mu_F^2}\right)-1\right)
\left(
y\ln(-y)-(y+1)\ln(y+1)
\right)
\right] &\nonumber\\
&
-\frac{V(z,y)}{y+z}+\frac{y\ln(-y)+(y+1)\ln(y+1)}{y(y+1)}
+\frac{y(y+1)+(y+z)^2}{(y+z)^2}W(z,y)
\Biggr\}+\Biggl\{z\rightarrow\bar z\Biggr\} \,  ,& \nonumber
\end{eqnarray}
where $\mu_R \, (\mu_F)$ is the normalization (factorization) scale,
$C_F=(N_c^2-1)/2N_c$, and
\begin{eqnarray}
&
V(z,y)=z\ln(-y)+\bar z\ln(y+1)+z\ln(z)+\bar z\ln(\bar z) &\nonumber \\
& W(z,y)=\Li_2(y+1)-\Li_2(-y)+\Li_2(z)-\Li_2(\bar z)+\ln(-y)\ln(\bar
z)-\ln(y+1)\ln(z) \, .
& \nonumber
\end{eqnarray}
The hard amplitudes for pion production in the
isoscalar state can be obtained by crossing from those for the isovector state.
They correspond to each other after the interchange
of the $t$-channel and the $s$-channel partonic pairs and the appropriate
interchange of the relative partonic momentum fractions. One can convince
oneself of this relationship easily by comparing the typical NLO diagrams
on the upper and lower line
in the Fig.~\ref{f1}. The prescription for the interchange of the momentum
fractions reads
    \begin{equation}
   z\leftrightarrow \frac{\xi+\tau}{2\xi}\, ,
   \quad \bar z \leftrightarrow \frac{\xi-\tau}{2\xi}\, ,
   \end{equation}
then
$Q^{(-)}\left(z,\tau\right)=Q^{(+)}\left(\frac{\xi+\tau}{2\xi},\xi(2z-1)\right)$,
and similar relations hold for $ R^{(-)},G^{(-)}$.

\section{Numerical results}
\begin{figure}[t]
\begin{center}
    \includegraphics[scale=1.]{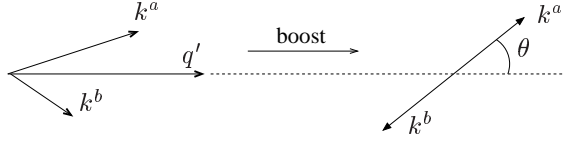}
\end{center}
    \caption
    {
     Definition of the polar angle $\theta$ in the c.m.\ of the
        pion pair.
     \label{theta}
    }
\end{figure}
We will present results for normalized Legendre moments (NLMs)
defined as a convolution of differential cross section
\begin{equation}
    \langle { P_n(\cos \theta)}\rangle^{\pi^a\pi^b}=
    \frac{
      \int\limits_{-1}^1 \mathrm{d}\cos \theta\,\;
      {P_n(\cos \theta)}\,
      \frac{\mathrm{d} \sigma^{\pi^a\pi^b}}{\mathrm{d} \cos \theta}
         }
         {
         \int\limits_{-1}^1 \mathrm{d}\cos \theta\,\;
         \frac{\mathrm{d} \sigma^{\pi^a\pi^b}}{\mathrm{d} \cos \theta}
         }
\end{equation}
with a Legendre polynomial $P_n(\cos \theta)$,
 where the polar angle $\theta$ is defined in
Fig.~\ref{theta}.

The GPDs in our calculation are modeled using the ansatz for double
distributions suggested by Radyushkin \cite{Radyushkin:1998es}.
Our models for 2$\pi$DAs reads
\begin{eqnarray}
 &  \Phi^{I=1}(z,\zeta,\mpp)=6z(1-z)(2\zeta-1)F_\pi(\mpp) \, , \\
 &  {  \Phi^{I=0}(z,\zeta,\mpp)=}   -\frac{120 M_2^Q}{n_f}z(1-z)(2z-1)
    \left[
    \frac{3-\beta^2}{12}f_0(\mpp)P_0(\cos \theta)-
    \frac{\beta^2}{6}f_2(\mpp)P_2(\cos\theta)
    \right] \, ,
   \nonumber  \\
&    {\Phi^G(z,\zeta,\mpp)= } -60 M_2^G z^2(1-z)^2
    \left[
    \frac{3-\beta^2}{12}f_0(\mpp)P_0(\cos\theta)-
    \frac{\beta^2}{6}f_2(\mpp)P_2(\cos\theta)
    \right] \nonumber \, ,
  \end{eqnarray}
where $2\zeta -1=\beta \cos\theta$, the pion velocity in the di-pion c.m.\
      is
$\beta=\sqrt{1-4m^2_\pi/m^2_{\pi\pi}}$,
and $n_f$ is the number of active flavors.  $M_2^Q$ and $M_2^G$
represent the momentum fractions carried by quarks and gluons in the pion target.
The isovector pair is produced in a $P$-wave, and $F_\pi(\mpp)$ is
the timelike electromagnetic pion form factor. The isoscalar pair may
be in an $S$- or $D$-waves, and $f_0(\mpp)$ and $f_2(\mpp)$ are the corresponding Omn\`es
functions for $S$- and $D$-waves.
Omn\`es functions develop an imaginary part above two-pion threshold. In the region where
pion-pion scattering is elastic, the phases of $f_0(\mpp)$ and $f_2(\mpp)$ coincide with the
pion phase shifts $\delta_0^{I=0}$, $\delta_2^{I=0}$ (Watson theorem). For more details
see \cite{Polyakov:1998ze,Lehmann-Dronke:1999aq}. For higher
di-pion masses the phases of Omn\`es functions do not coincide
with the pion phase shifts. In the present analysis we neglect
inelasticity and use the pion phase
shifts
as an input in dispersion relations to reconstruct the Omn\`es
functions.
We
use two sets (S1 and S2) of
parameterizations  for the Omn\`es functions. In the set S1 we calculate Omn\`es
functions using dispersion relations with two subtractions and the
fits  \cite{Pelaez:2004vs} for the pion phase shifts. For the subtraction constant we
used the result of \cite{Polyakov:1998ze}.
The set S2 is the same as that one used in
\cite{Lehmann-Dronke:2000xq}.

Odd Legendre moments are proportional to the
product of isoscalar and isovector amplitudes
\begin{equation}
\langle P_{1,3}\rangle \propto \int d\cos\theta \,\, P_{1,3}
(\cos\theta)\, Re\left\{\left(T^{I=1}\right)^*T^{I=0}\right\}
\end{equation}
and sensitive to the interference of $P$-wave with $S$- or $D$-
wave,
\begin{equation}
\langle P_1\rangle \propto Re\left\{F^*_\pi(\mpp)(c_1 f_0(\mpp)+c_2 f_2(\mpp))\right\}
\, , \quad
\langle P_3\rangle \propto Re\left\{F^*_\pi(\mpp)f_2(\mpp)\right\}
\, .
\end{equation}
These observables  provide access to a small isoscalar
amplitude.
\begin{figure}
\begin{center}
{\includegraphics[height=0.427\textheight,width=0.9\textwidth,clip]{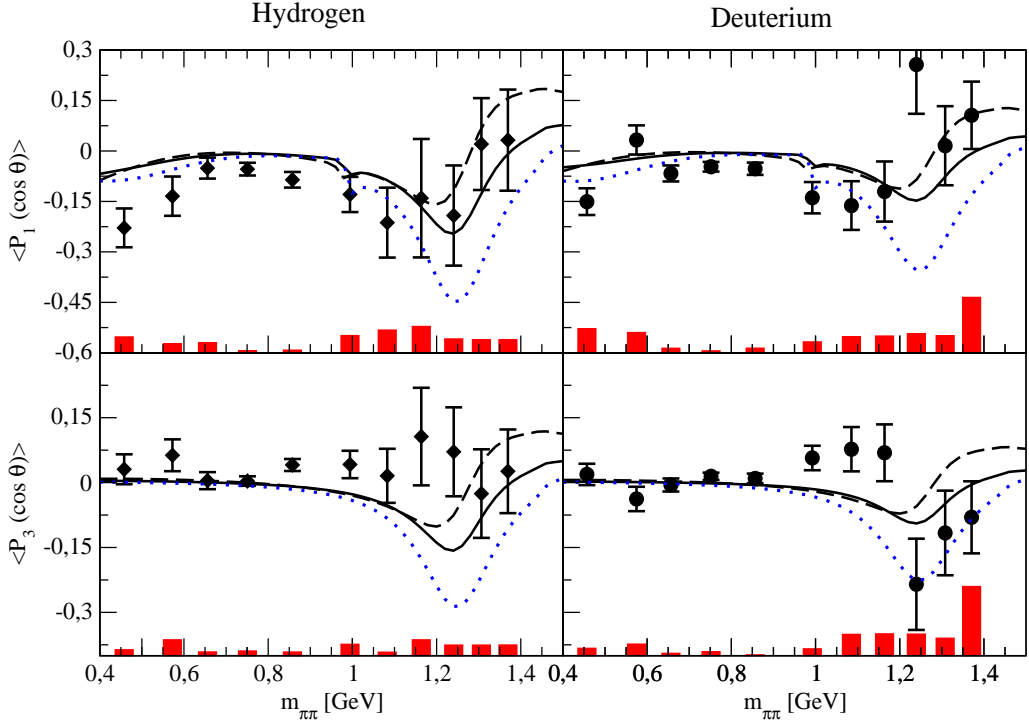}}
\end{center}
\caption{
The $m_{\pi\pi}$ dependence
of the
normalized Legendre moments $\langle P_1\rangle$ (upper panels) and
$\langle P_3\rangle$ (lower panels) for hydrogen (left panels) and
deuterium
 (right panels).  The curves show our
results at LO (dashed black), NLO (solid black) and at NLO without
two-gluon exchange in the $t$-channel (dotted).  The average
kinematics for the HERMES data \protect\cite{Airapetian:2004sy} is
$\langle x_\mathrm{Bj} \rangle=0.16$,
$\langle Q^2\rangle=3.2\,(3.3)\,\textrm{GeV}^2$, and $\langle
-t\rangle=0.43\,(0.29)\,\textrm{GeV}^2$ for hydrogen (deuterium).
The histograms show the
systematic uncertainty of the measurement.}
\label{hermes}
\end{figure}

In Fig.~\ref{hermes} we compare the results of our calculation (with set S1)
with HERMES data for NLMs on hydrogen and deuterium targets
\cite{Airapetian:2004sy}.
Note that the difference between the LO and NLO results
is not big, which may indicate a fast convergence of the
perturbative series. Our leading twist predictions for $P_1$
are in reasonable agreement with the experimental data
for the hydrogen and deuterium targets.
Since such experiments
in principle allow one to
test the gluonic
content of the nucleon, we plot in addition the corresponding
results when we omit the two-gluon exchange in the $t$-channel. The
effect of the gluon contribution is noticeable, hence improved
data in the future may
provide constraints on
the gluonic content of
the nucleon.

In Fig.~\ref{compas} we present our results
for kinematics typical of the COMPASS experiment, i.e.\ for
      larger $Q^2 $ and smaller
$x_{\rm{Bj}}$.
We plot here the results obtained with
different Omn\`es functions, with and without account of two-gluon
exchange the $t$-channel. The predicted
values of the NLMs are smaller for COMPASS kinematics than
that ones for HERMES. Another
observation is that the gluon GPD plays here even a more important
role.
\begin{figure}
\begin{center}
\subfigure{\includegraphics[width=0.85\textwidth,clip]{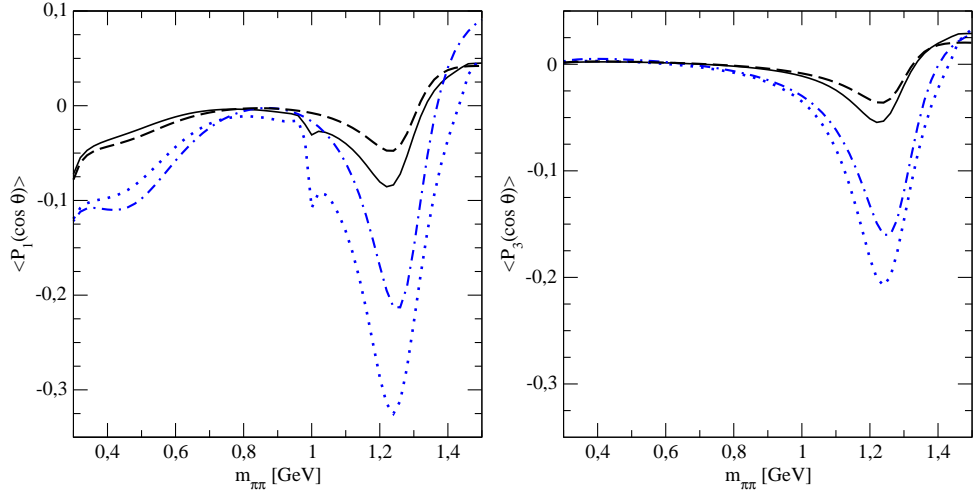}}
\end{center}
\caption{The normalized Legendre moments
calculated for a deuterium target in the kinematic region relevant
for COMPASS ( $x_\mathrm{Bj}=0.08$, $-t=0.27\,\textrm{GeV}^2$ and
$Q^2=7\,\textrm{GeV}^2$)
with sets S1 (solid) and S2 (dashed lines). The dotted and dashed-dotted lines represent the
calculation without two-gluon exchange in the $t$-channel, for sets S1
and S2 respectively.}
\label{compas}
\end{figure}

To estimate the scale uncertainties of our NLO calculation we plot
in Fig.~\ref{nlo} for HERMES kinematics the
"longitudinal" combination of the moments, which projects out the state
with vanishing total helicity of the pion pair, calculated with set S1
and different settings in the hard part for the factorization and
the renormalization scales. We see that this
scale uncertainty is not large.
\begin{figure}
\begin{center}
\includegraphics[width=0.65\textwidth,clip]{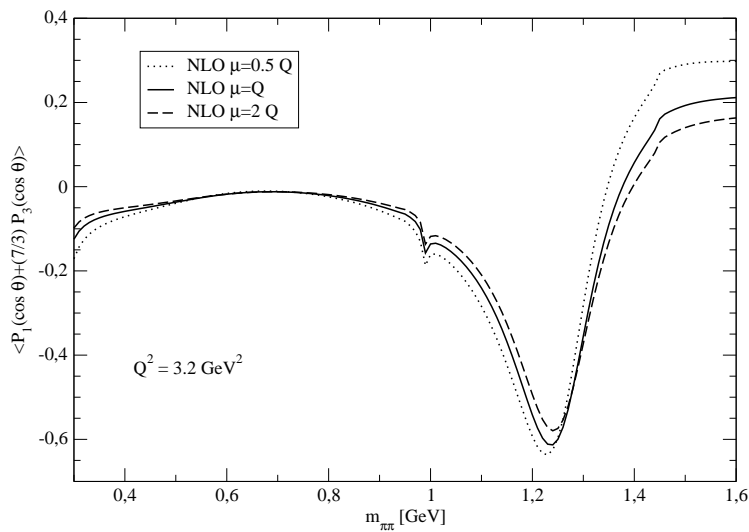}
\end{center}
\caption{The "longitudinal" NLM combination calculated  for
different renormalization and factorization scales,
$\mu_R=\mu_F=\mu$. }
\label{nlo}
\end{figure}

\section{Summary}

We studied di-pion electroproduction in the QCD factorization
approach at NLO.
Numerical results obtained for HERMES and COMPASS kinematics
show that NLO corrections (at least for normalized Legendre
moments) are small.
Hence with improved
experimental data from such experiments it will
be possible to put the constrains on both GPDs and 2$\pi$DAs. One can also get
additional insight into the gluonic content of the nucleon. To be
on the safe side with the leading twist approach it is, however,
important to go to larger $Q^2$.

In the future we want to improve our approach to Omn\`es
functions by taking into account inelasticity effects.

\end{document}